\documentclass{llncs}

\usepackage{amssymb}
\usepackage{color}
\usepackage{amsmath}
\usepackage{graphicx}
\usepackage{caption}
\usepackage{subcaption}
\PassOptionsToPackage{hyphens}{url}\usepackage{hyperref}
\usepackage{enumitem}

\definecolor{orange}{rgb}{1,0.5,0}

\newcommand{\done}[1]{}

\begin{document}

\title{FEEBO: An Empirical Evaluation Framework for Malware Behavior Obfuscation}

\author{Sebastian Banescu \and Tobias W\"uchner \and Marius Guggenmos \and Mart\'in Ochoa \and Alexander Pretschner}
\institute{Technische Universit\"at M\"unchen,
 Germany\\
}

\maketitle

\begin{abstract}
Program obfuscation is increasingly popular among malware creators. Objectively comparing different malware 
detection approaches with respect to their resilience against obfuscation is
challenging. To the best of our knowledge, there is no common empirical 
framework for evaluating the resilience of malware detection approaches w.r.t. behavior obfuscation.
We propose and implement such a framework that obfuscates the observable behavior of malware binaries. 
To assess the framework's utility, we use it to obfuscate 
known malware binaries and then investigate the impact on detection 
effectiveness of different $n$-gram based detection approaches.
We find that the obfuscation transformations employed by FEEBO significantly affect the precision of such detection approaches.	Several $n$-gram-based approaches can hence be concluded not to be resilient against this simple kind of obfuscation.
\end{abstract}

\section{Introduction}
\done{TW: consolidate terms: event vs. call; profile vs. configuration vs. setting}
Malware continues to be a relevant cyber security threat. While in the early days of the Internet malware was often developed for the pure sake of curiosity, malware development today follows a clear-cut business model. The motivations to develop and utilize malware ranges from supporting cyber espionage over theft of confidential data, denial-of-service of commercial services, or even black-mailing, up to tampering with military or civilian infrastructures.

Industry and academia continuously devise countermeasures to cope with this threat in form of advanced malware detection approaches. However, malware developers are often several steps ahead the state of the art. 
Most commercial antivirus software in principle continues to be 
some form of signature-based analysis on the persistent representation of potential malware. Not surprisingly, almost all modern malware families employ some means to confuse and hamper signature-based approaches. Such countermeasures range from simple techniques (e.g.~build-time encryption and runtime decryption up), to more sophisticated techniques (e.g. control-flow obfuscation or anti-debugging mutations) \cite{Collberg2009}. 

Given control-flow obfuscation of today's malware, one intuitively appropriate detection strategy is so-called behavioral detection. The idea is to look at the malware's runtime behavior rather than its static code. This behavior includes issued function or system calls, or in general, every runtime interaction with system resources. By construction, behavioral detection approaches are barely affected by control-flow obfuscation. 
However, although behavioral detection techniques compensate the effects of (build-time) control-flow obfuscation techniques to a large extent, they are often vulnerable to more advanced (run-time) behavior obfuscation techniques that ``blur'' the externally visible behavior of malware. 
Examples for such behavior obfuscation techniques include the injection of bogus system calls or the deliberate randomized re-ordering of call execution sequences.

While control-flow obfuscation of malware and respective 
countermeasures at the detection side have been well 
researched~\cite{Collberg2009}, the effects of 
\emph{behavior} obfuscation on the effectiveness of detection approaches so 
far only received very little attention in the literature. Behavior obfuscation 
in itself has been discussed from a theoretical 
perspective~\cite{dalla2008semantics}, but we are not aware of any empirical investigations of the effects of behavior obfuscation of real-world malware.

To provide a foundation for such empirical evaluations, we propose a behavior obfuscation framework which we call FEEBO. Provided an arbitrary malware sample as input, it applies a diverse set of behavior obfuscation transformations to its externally visible behavior, which is defined by issued system calls. This makes it possible to ``inject'' behavior obfuscation mechanisms into malware samples in a structured and targeted way, regardless of whether or not the specific malware sample performs any behavior obfuscation itself. Considering that behavior obfuscation at the system call level is still rarely done by real-world malware, this approach allows us to get one step ahead of malware developers and reason about the impact of such obfuscation techniques on state of the art detection approaches before they are implemented and released into the wild. 

\textbf{Contributions}: \textbf{a)} To our best knowledge, we are the first to propose an empirical malware behavior obfuscation framework that is able to behaviorally obfuscate standard malware binaries. \textbf{b)} With FEEBO we establish a basis for a wide range of reproducible behavioral obfuscation resilience experiments. \textbf{c)} Our evaluations show that for certain configurations, the precision of $n$-gram ~\cite{wressnegger2013} based detection approaches are significantly affected by behavioral obfuscation.

\textbf{Organization}: We introduce the concept of behavior obfuscation and discuss two main representative $n$-gram-based behavioral detection techniques in \S\ref{sec:preliminaries}. Then we describe the design and implementation of FEEBO in \S\ref{sec:approach}. We show the effectiveness of a prototypical implementation of our framework and discuss its limitations in \S\ref{sec:evaluation}. We discuss possible application areas of our approach and give an outlook on future work in \S\ref{sec:conclusion}.

\section{Preliminaries}
\label{sec:preliminaries}

We start with some relevant concepts from the literature. In particular we recall related work
on behavior obfuscation and detection based on $n$-grams.

\subsection{Behavior Obfuscation}

This paper is inspired by the work of P\'{e}choux and Ta~\cite{pechoux2014} on behavior obfuscation of malware. They divide the behavior, i.e.~executed operations of a program (e.g.~malware) into (i) internal computations and (ii) system calls. \emph{Internal computations} operate only on the process memory of the corresponding program and they only affect and are affected by the information stored inside this process' memory. \emph{System calls} represent interactions with the operating system (OS) kernel, i.e.~there is a transfer of control from the corresponding program to the kernel and back. Therefore, system calls affect and are affected by the information stored anywhere in the OS memory. 

The sequence of system calls performed by a program is called the \emph{observable (execution) path} or \emph{behavior}. P\'{e}choux and Ta show that it is possible to transform (obfuscate) the observable path of known malware samples such that the original malware functionality is preserved by: (i) inserting system calls before and/or after system calls in the observable path, (ii) reordering system calls in the observable path and (iii) substitution of system calls by other system calls which provide at least the same functionality. Different from our work, their goal is to obtain a trace that is similar to a goodware trace (mimicry). We, on the other hand, focus on randomly generating sets of malware ``mutants'' to assess their effect on behavioral detection approaches that analyze the system calls executed by malware.

There is an important difference between \emph{behavior obfuscation} and \emph{control-flow obfuscation}. Control-flow obfuscation applies transformations at the source code or intermediate representation levels in order to make a program harder to understand by a human or an automated analysis engine. Such code transformations include virtualization obfuscation, insertion of bogus code via opaque predicates, function splitting, and control-flow flattening \cite{Collberg2009}. These transformation will typically not have an effect on the observable execution path of that program. On the other hand, behavior obfuscation strictly implies changing the observable execution path of the program being obfuscated.


\subsection{Behavioral Malware Detection}

In contrast to approaches that focus on the persistent representation of malware, behavioral detection approaches discriminate malware from goodware by establishing characteristic behavior profiles. 
Such approaches range from using raw system call traces to short sequences of 
calls, so-called $n$-grams~\cite{Forrest1996,Rieck2011,wressnegger2013}, to more elaborate concepts that model the semantic interdependencies between different calls in call-graphs~\cite{Park2013,Christodorescu2005,Christodorescu2008}. There also exist approaches that model behavior by abstracting system calls into induced data flows~\cite{Bhatkar2006,Cavallaro2011}. These approaches are based on traces of issued system calls and are thus likely to be affected by the aforementioned behavior obfuscation transformations.

In this study we focus on approaches that base on $n$-grams as a behavior model, due to its prevalence in academic publications~\cite{wressnegger2013}. We are aware that findings based on this model do not necessarily generalize. Nevertheless we are convinced that such an evaluation is a good starting point to reason about the effects of behavior obfuscation in general and will be the basis for future work.
To cover a broad range of $n$-gram based detection approaches, we follow the categorization schema of Canali et al.~\cite{canali2012quantitative}. We consider $n$-grams built on system calls without arguments as atoms and both a) considering or b) ignoring the ordering of calls for their construction.
To test the aforementioned approaches we first executed known malware and goodware in a sandboxed environment and monitored their executed system calls. This procedure yielded labeled event logs, which we tokenized with a sliding window, moving a window of defined but fixed size over the respective log, thus yielding sets of $n$-grams of system calls.

For the first $n$-gram approach, which considers the ordering of system calls (a), we directly feed the obtained $n$-grams as features into a supervised machine learning classifier. 
For the second $n$-gram approach (b), which does not consider the ordering of system calls (b), we count the number of occurrences of each system call in the $n$-gram, build a feature vector with the number of occurrences of each of the system calls in the $n$-gram, and feed these vectors into the classifier.
Note the independence of the feature vectors from the ordering of system calls in the $n$-gram. 

\begin{figure}[t]
\begin{center}
	\vspace{-1em}
		\includegraphics[width=0.8\textwidth]{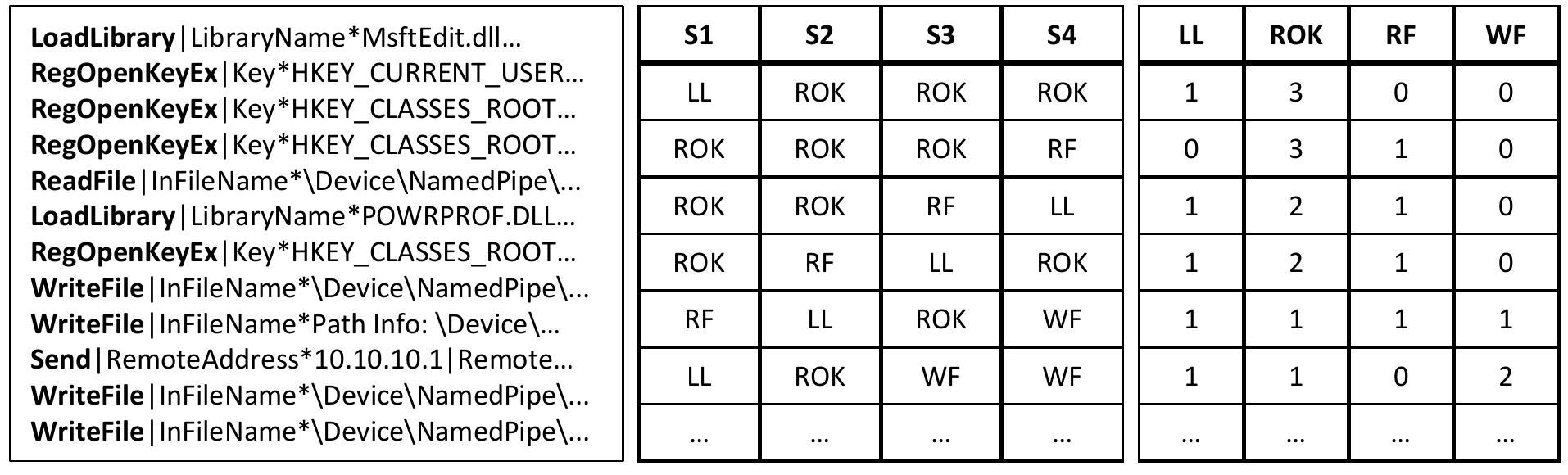}\vspace{-0.5em}
	\caption{Call trace vs. ordered $n$-gram (a) vs. unordered $n$-gram (b)} \vspace{-2em}
	\label{fig:ngram}
\end{center}
\end{figure}

Figure~\ref{fig:ngram} depicts the resulting feature vectors for both approaches when applied to a small sample call trace (left). The middle shows n-grams for approach (a), consisting of 4 system calls on each row (i.e, 4-grams). The contents of the cells are the initials of the system calls from the trace to the left. The table on the right part shows $n$-grams for approach (b), which consist of the frequency of every system call (depicted in the table header) for a 4-gram on each row. 

\section{Our Approach to Behavior Obfuscation}
\label{sec:approach}

Transforming (obfuscating) x86 binary programs without debugging symbols is a non-trivial task which involves binary rewriting \cite{prasad2003binary}. This task becomes even more challenging when the binaries we want to transform are malware, which employ anti-disassembly techniques \cite{eagle2011ida}. However, since we only want the binary to have a different observable behavior in terms of systems calls, we have taken an alternative approach by using binary instrumentation \cite{reddi2004pin}.

In a nutshell binary instrumentation allows one to intercept any system calls performed by the target binary. One can choose to execute, delay, drop or even swap the intercepted system call, plus perform other additional instructions including making more system calls. 
We have implemented the following two behavior obfuscation transformations: (i) system call \emph{insertion} and (ii) system call \emph{reordering}. These are relatively simple techniques in comparison to substitution of system calls with functionally equivalent systems calls; we leave their implementation to future work. 

\subsection{System Call Insertion} \label{subsec:insertion}
With a given probability $p_i$, system call insertion adds for each system call made by the obfuscated application a number of additional system calls randomly chosen from the previously executed system calls. The number of inserted system calls is randomly chosen between $\mathit{min_i}$ and $\mathit{max_i}$, two more input parameters of FEEBO. To prevent these inserted calls from changing the original functionality of the application, we modify the values of their parameters in case the system calls belong to a set $S$ of calls that have side-effects such as writing to a file. The values of the changed parameters are chosen such that they will not collide with existing data, e.g., files. Furthermore, system calls that access a unique system resource are excluded. For instance, if we were to insert the system call that sets the clipboard data, we would need to also insert a second call to restore the clipboard data since there is only one clipboard on each system. 

For example, with $p_i = 0.25$, $\mathit{min_i} = 2$ and $\mathit{max_i} = 5$, every system call made by the application has a 25\% chance to insert a randomly chosen number between 2 and 5 of system calls after the execution of the intercepted system call. This obfuscation transformation changes the externally visible behavior by inserting a random number of system calls in random locations of the original execution trace. The intuition is that it should be effective against $n$-grams-based detection approaches since they rely on patterns.

\subsection{System Call Reordering} \label{subsec:reordering} 
System call reordering can na\"{i}vely be implemented by delaying a sequence of system calls in a buffer which is randomly permuted before execution. This would most likely break the functionality of the transformed program or even cause it to crash. Instead, every system call in $S$ executed by the transformed application, can be delayed with probability $p_r$ and placed in a queue (of size $n$) for later execution. The reason only calls in $S$ are being delayed is that calls outside $S$ generally read information which applications need to continue their proper execution. Moreover, we use a queue for the delayed system calls, because we want to preserve the original ordering of system calls that have side effects like writing to a file. Once the queue reaches a certain size, our tool will execute them in their original order.
Each of the delayed calls can additionally trigger the insertion other system calls with similar parameters as described in \S\ref{subsec:insertion}, i.e. probability of insertion denoted $p_{ri}$ and the minimum and maximum number of inserted system calls, denoted by $\mathit{min}_{ri}$, respectively $\mathit{max}_{ri}$. 

For example, for $p_r = 0.5$, $n = 5$, $p_{ri} = 0.75$, $\mathit{min}_{ri} = 1$ and $\mathit{max}_{ri} = 2$, every system call from $S$ made by the application is delayed with a 50\% probability. Once 5 calls have been delayed, they will be executed. Each of the delayed executions has a 75\% probability to insert one or two other system calls.

\begin{table}[t]
 \centering
 \begin{tabular}{|l | c | c | c | c | c |}
  \hline
  & $p_{\{i,ri\}}$ & $min_{\{i,ri\}}$ & $max_{\{i,ri\}}$ & $p_r$ & $n$ \\
  \hline \hline
  System Call Insertion & $[0,1]$ & $\{0, \dots, max_{\{i,ri\}}\}$ & $\{ min_{\{i,ri\}}, \dots, \infty\}$ & -- & -- \\
  \hline
  System Call Reordering & $[0,1]$ & $\{0, \dots, max_{\{i,ri\}}\}$ & $\{ min_{\{i,ri\}}, \dots, \infty\}$ & $[0,1]$ & $\{ 0, \dots, \infty\}$ \\
  \hline
 \end{tabular}
 \vspace{0.5em}
 \caption{Obfuscation transformations versus parameters}
 \label{tab:params}\vspace{-1.75em}
\end{table}

\subsection{Obfuscation Profiles} \label{subsec:profiles}
The range of the input parameters of the previously described obfuscation transformations are shown in Table~\ref{tab:params}. The insertion and reordering probabilities range from 0 to 1. The minimum and maximum numbers of inserted system calls as well as the size of the reordering queue are positive integers. Their upper bound depends on the data type and the architecture of the system they are running on.
Based on these parameters of system call insertion and system call reordering we can configure various obfuscation profiles, e.g.~``always insert 2 system calls after each system call in the original observable path'', ``do not insert any calls, only reorder'' or ``insert 1 system call after each reordered call''. We will see concrete detection values for different obfuscation profiles in \S\ref{sec:evaluation}.

\section{Evaluation}
\label{sec:evaluation}

To assess the applicability of FEEBO, we obfuscated a set of real-world malware with the help of FEEBO and then applied the previously introduced behavior detection approaches, based on n-grams of system calls, to the resulting obfuscated system call traces.

\paragraph{Setup.}
We executed 100 malware samples within an installation of the Cuckoo malware analysis sandbox\footnote{\url{http://www.cuckoosandbox.org/}.}, where we replaced the behavior monitor with FEEBO to obtain a variety of obfuscated behavior traces of those samples. In addition, we collected the traces of 100 known goodware samples which we did not obfuscate, to use as comparison baseline for later training the detection classifiers. 

The large range of values that the obfuscation parameters can take (see Table~\ref{tab:params}, quickly leads to a combinatorial explosion of the obfuscation profiles. Moreover, to capture a critical mass of system calls sufficiently large to allow training a classifier with good accuracy, we need to monitor a malware sample for at least 3 minutes. With one configuration profile capturing the obfuscated traces of 100 malware samples would then take 300 minutes which, with help of parallel execution of multiple VMs on 5 cores, we could cut down to about one hour per run. Therefore, we conducted experiments with 375 different combinations of the obfuscation parameters. More specifically, we set all probabilistic parameters like the insertion or reordering probability to selected values between 0\% and 100\%, i.e. $p_{\left\{i,r,ri\right\}}\in\left\{0.0,0.25,0.5,0.75,1.0\right\}$, and particular interesting discrete parameters to fixed values between 1 and 10, i.e. $max_i\in\left\{1,5,10\right\}$. All other parameters were set to fixed values, i.e.~$min_{\{i,ri\}} = 1$, $max_{ri} = 3$ and $n = 5$. 
Conducting one evaluation for each configuration profile (e.g. one combination of the aforementioned parameters and value ranges) ends up in $5\times5\times5\times3=375$ runs, which sums up to a total runtime of about 16 days. 

\paragraph{Experiments.}
Using the resulting execution traces, we trained the respective classifiers on the feature vectors computed on the non-obfuscated baseline traces and used the generated classifier on the remaining obfuscated event traces. For the ideal case of the applied obfuscations not having any effect on the externally visible behavior, the detection rate should remain 100\%.
With this setting we could investigate the effects of the applied obfuscation transformations with respect to detection accuracy. To assess the effects of different $n$-gram sizes we repeated this procedure for all possible $n$-grams for $n$ between $3$ and $10$.

\begin{figure}[t!]
		\vspace{-2em}
				\begin{center}
         \includegraphics[width=1\textwidth]{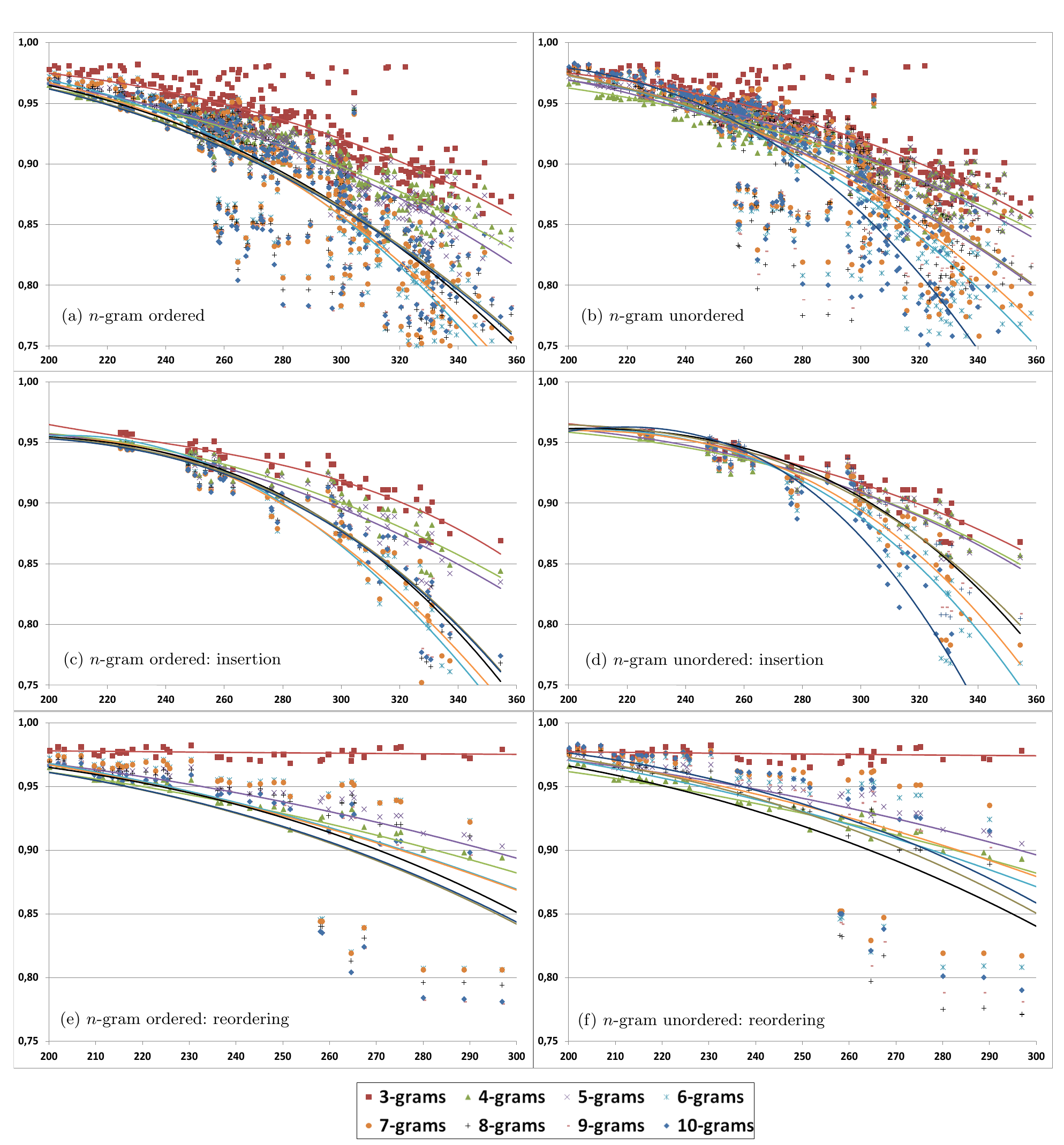}

        \caption{Degree of Obfuscation ($x$-axis) versus Detection Rate ($y$-axis)}

				\label{fig:obfuscation_results}
				\end{center}
\end{figure}

Figure~\ref{fig:obfuscation_results} summarizes the experimental findings. As a measure of the degree of obfuscation, we calculated the \emph{Levenshtein distance} between the respective traces, as it represents the number of atomic insertion, deletion, and substitution operations that are needed to transform one event trace into another one.
For computing the Levenshtein distance we abstracted our traces to only the name of the system calls (not their parameters), which are elements of our alphabet.
Correspondingly, the $x$-axis of each diagram represents the average obfuscation degree of all considered event traces, whereas the $y$-axis represents the detection rate (percentage of correctly identified malware samples) achieved by different detection approaches. To visualize the development of the median detection rate for increasing obfuscation degree we also plot trend-lines for each $n$-gram.

We split the evaluation results into three parts: the first row represents the results for the experiments where both type of obfuscation transformations, i.e.~call reordering and call insertion were applied; the second row illustrates the results for the insertion experiments; and the last row the results from the reordering experiments.
As we can see, the applied obfuscation transformations have a significant effect on the detection effectiveness of the $n$-gram approaches. In the first row of Figure~\ref{fig:obfuscation_results} we can deduce a roughly quadratic relationship between an increase of obfuscation degree and an decrease of the detection rate. Also we can see, that the spread in classification accuracy, i.e. the standard deviation of the detection rate, significantly rises the more obfuscation is applied. Furthermore we can see that higher-order $n$-grams are more sensitive towards obfuscation.

Looking at the remaining diagrams we notice that insertion transformations seem to have a bigger impact on detection accuracy than reordering transformation, which is reflected in a significantly steeper slope of the trend-lines in the insertion diagrams than in the reordering diagrams.
Also we can say that for very small $n$-gram sizes, reordering transformations seem to have barely any influence on the detection rate, as can be seen by almost constantly high detection rates. 
Finally, our evaluations did not reveal any significant difference in obfuscation resilience between the ordered and unordered types of $n$-gram approaches.

\paragraph{Discussion and threats to validity.}
First note that although we only conducted one execution run for one constellation 
of configuration parameter, the fact that several parameter configurations 
lead to a sample with a similar Levenshtein distance allows us to achieve a 
good saturation of the obfuscation spectrum. Given that we obtain 375 distinct 
sets of 100 obfuscated event traces, for each profile in our experiment, this gives a 
rather high density of 41 data-points in a range of 20 units on the $x$-axis in the first row from Figure~\ref{fig:obfuscation_results}, which correspond to the ``both insertion and reordering'' obfuscation profile. However, the density is 7 data-points in a range of 20 units for the second and third rows which correspond to ``insertion-only'', respectively ``reordering-only'' obfuscation profiles.
\done{TW: make point more clear that we did not need more runs as we were mainly interested in the obfuscation effects and not so much in the causing parameters and the same effect could be achieved by a variety of different configuration settings}


We intentionally did not mention false positive rates of the $n$-gram approaches in our evaluation, because they are not relevant for our experiments, since we do not change or obfuscate the set of goodware during our experiments. 
Currently our experimental setting assumes the presence of a certain ground truth, i.e. the availability of a critical mass of unobfuscated malware for classifier training. If malware developers start to make more use of behavioral obfuscation mechanism the availability of such a basic training set is not guaranteed. Using obfuscated malware for both, testing and training the classifiers, will likely diminish their effectiveness even more. For future work, we therefore also plan on investigating whether these factors impact the results. 

Having performed some initial experiments with Na{\"i}ve Bayes, Gaussian-kernel SVMs, and Random Forest classifiers, we can confirm that the choice of the baseline classifier does not have a significant effect on the relative obfuscation sensitivity of the considered $n$-gram approaches.

The functionality of any obfuscated program should include the functionality of the original (non-obfuscated) program. For many software transformation engines such as optimizing compilers, this is a strict requirement. However, even very widely used compilers such as GCC or Clang have been found to contain optimizations that break the functionality of the original source code~\cite{Yang2011}. The \emph{system call reordering} transformation described above suffers from the same issue, i.e.~it may change the functionality of malware such that it becomes ineffective. 
Arguably, however, in the case of obfuscating widely used malware it is more important to avoid detection even if the obfuscation engine will output some samples which are not effective. We do not yet possess statistics regarding the number of effective malware samples output by our tool. However we plan to study this fact as part of future work. We still consider our results valuable given that checking for behavioral equality in general is not decidable and in our experiments none of the obfuscated malware samples crashed during execution.

In sum, we can draw two main conclusions from our experiments: a) FEEBO is able to effectively obfuscate the behavior of real-world malware with significant effect on the effectiveness of behavioral detection approaches; b) the considered type of $n$-gram approaches is highly sensitive to the evaluated forms of behavior obfuscation.

\section{Conclusions and Future Work}
\label{sec:conclusion}

We have introduced FEEBO, a framework to conduct empirical experiments on the effects of behavior obfuscation on malware detections. To this extent we developed a prototype that can apply certain obfuscation transformations to the externally visible behavior of malware samples. 
To evaluate the effectiveness of the implemented obfuscation transformations and of our approach in general, we investigated the effects of a wide range of behavior obfuscation transformations on the detection capabilities of two representative $n$-gram behavior detection approaches. 
We could show that both types of $n$-gram approaches are considerably vulnerable to the applied obfuscation transformations. 

We are aware that our presented evaluation results are not comprehensive 
in its present form. In particular, for future work we plan to repeat the experiments for a bigger configuration space and malware sets. We also plan to investigate the effects of lack of ground truth by training the classifiers on obfuscated malware samples instead on solely unobfuscated ones. 
In terms of possible extensions of FEEBO, we plan to implement additional obfuscation transformations that e.g. also tackle the substitution of certain system calls with semantically equivalent ones. 

Although we release FEEBO\footnote{\url{https://www22.in.tum.de/tools/feebo/}} to parties from academia and industry, for ethical reasons we will provide a version that is not capable of generating self-contained obfuscated malware binaries. Instead, FEEBO needs to be manually installed in the evaluation environment, together with a installation of \emph{Intel Pin} \cite{reddi2004pin}, which hopefully hampers misuse of FEEBO by malware developers.

\bibliographystyle{abbrv}
\bibliography{bibliography} 

\end{document}